\documentclass[12pt]{article}
\usepackage[utf8]{inputenc}
\usepackage{graphicx}
\usepackage[utf8]{inputenc}
\usepackage{amsmath}
\usepackage{amsfonts}
\usepackage{amssymb}
\usepackage{graphicx}
\usepackage{setspace}
\usepackage{authblk}
\usepackage{caption}
\usepackage{float}
\usepackage{cite}
\usepackage[left=2cm,right=2cm,top=2cm,bottom=2cm]{geometry}
\usepackage{color}
\usepackage{color,soul}
\title{\textbf{Reconstruction of DBI-essence dark energy with $f(R)$ gravity and its effect on black hole and wormhole mass accretion}}
\author[1]{Alokananda Kar}
\author[2]{Shouvik Sadhukhan}
\author[3]{Ujjal Debnath}
\affil[1]{\small{Department of Physics, University of Calcutta,
Kolkata-700073, West Bengal, India. \par
Email:{alokanandakar@gmail.com}}} \affil[2]{\small{Department of
Physics, Indian Institute of Technology, Kharagpur-721302, West
Bengal, India. \par Email: shouvikphysics1996@gmail.com}}
\affil[3]{\small{Department of Mathematics, Indian Institute of
Engineering Science and Technology, Shibpur, Howrah-711 103, West
Bengal, India.\par Email: ujjaldebnath@gmail.com}}

\date{}
\begin{document}
\maketitle

\begin{abstract}
In this paper, we have used the reconstructed DBI-essence dark
energy density to modify the mass accretions of black holes and
wormholes. In general, the black hole mass accretion does not
depend on the metric or local Einstein geometry. That is why we
have used a generalized mechanism by reconstructing the
DBI-essence dark energy reconstruction with $f(R)$ gravity. We
have used some particular forms of the scale factor to analyze the
accretion phenomena. We have shown the effect of cosmic evolution
in the proper time variation of black hole mass accretion.
Finally, we have studied the validity of energy conditions and
analyzed the type I - IV singularities for our reconstructed
model. \\

\textbf{Keyword}: black hole mass accretion, DBI-essence dark
energy, Reconstruction mechanism, $f(R)$ gravity, Cosmic evolution
\end{abstract}

\section{Introduction}

The black hole is one of the most important results of local
solutions of the Einstein equation. The evolution of the universe
follows the global solution. Although there is no direct
connection between the global and local solution in geometry, the
mass accretion can create a link between cosmic fluid and the
black hole or wormhole like local solutions.\par
The nature of cosmic phases depends on the cosmic
fluid with which it evolves. Various modified theories have been
proposed to discuss the different cosmic phases. The Einstein
Hilbert action fails to discuss the accelerated expansion of the
universe, which has been experimentally proven. The introduction
of the modified function of Ricci scalar in Einstein action can
provide the negative pressure, which in turn can give a repulsion
effect. This negativity can explain inflationary solutions as well
as the late-time acceleration. From the Raychaudhuri equation, it
is observed that the energy conditions must have positive values
for a gravitating system. However, positive energy conditions can
never produce accelerated expansion in cosmology. Therefore,
simple Einstein gravity geometry alone cannot discuss the
accelerated expansion properly. Modified gravity models can
resolve this problem with geometry modification. There are several
types of modified gravity available in the literature. Most useful
modifications are $f(R)$ gravity, Gauss-Bonnet gravity or $f(G)$
gravity, Teleparallel gravity or $f(T)$ gravity, non-Metricity of
gravity or $f(Q)$ gravity, and so on. $f(Q)$ and $f(T)$ come from
the modifications in the metric tensor and Christoffel
connections. $f(R)$ and $f(G)$ gravity models are basically the
higher-order modification of Ricci tensor in Einstein action
\cite{1,2,3,4,5,6,7,8,9,10,11,12}.\par The accelerated expansion
or expanding universe can also be explained with the exotic nature
of matter and energy. The most highlighted exotic models are the
dynamic dark energy models. There are three dark energy models,
namely fluid, scalar field, and Holographic models. The primary
requirement behind the use of these dynamics dark energy models is
to resolve the vacuum energy problem or cosmic coincidence problem
\cite{7,17,18,19,20,22}. \par Dirac-Born-Infeld
(DBI)-essence dark energy model \cite{13,14,15,16,17} is one of
the scalar field dynamic dark energy models that are used in
cosmic physics, even in string theory. There are several types of
scalar field models available in the literature, viz. Quintessence
model, K-essence model, Phantom field model, H-essence model, and
Tachyon model. The Tachyon model is one of the most generalized
scalar field models that can discuss all the cosmic phases and
conditions, but it fails to bring the cosmic Brane Bulk tension.
We have used this DBI-essence model of dark energy is used in our
study.\par All these cosmic models modify the activity of cosmic
fluids, and these modifications, in turn, modify the rate of mass
accretion of black holes. It was observed that the mass accretion
of the black hole and wormhole strictly depends on the cosmic
phases and their evolutionary scheme. Babichev et al. \cite{38,32}
discussed the accretion of phantom DE into a Schwarzschild black
hole, using the Holographic technique, and discussed that the mass
of the black hole would gradually decrease to zero near the big
rip singularity. Modified gravity modifies the nature of the
cosmic fluid, and that is how it can also modify the rate of mass
accretion of Black holes and wormholes. The reconstruction
mechanism can reproduce the evolution of energy density and
pressure of the cosmic
system.\cite{24,25,26,27,28,29,30,31,32,33,34,35,36,37}\par Here
in this paper, we have first reconstructed the energy density and
pressure of the DBI-essence Dark energy model with $f(R)$ gravity
and discussed the energy conditions of the system. This
modification has been used to discuss the black hole and wormhole
mass accretion. We have discussed the rate of mass accretion in
the presence of different scale factors and compared the rate with
different cosmic phases. We have also tried to provide some
physical reason behind the nature of mass accretion.\par So, the
paper is structured as follows.\par In sections 2 and 3, we have
discussed the basics of the modified gravity and scalar field
theory. Section 4 contains the coupling calculations that help us
to discuss the reconstruction formalism. Sections 5 and 6 have the
basics of thermodynamic energy conditions and condensed body mass
accretion. In section 7, we have discussed the analysis for
different scale factors and the comparison of their results. In
section 8, we have finally tried to resolve the finite-time future
singularities w.r.t. the scale factors, energy densities, and
pressures. We have provided the details of the
results in the discussion section 9 and the overall outcome of the
work in section 10.

\section{Overview of $f(R)$ gravity}

The study of $f(R)$ gravity should start from its
action as follows \cite{56}
\begin{equation}
    S=\frac{1}{16\pi}\int{f(R)\sqrt{-g}d^4x+}\int{L_m\sqrt{-g}d^4x}.
\end{equation}
Here $f(R)$ is a function of the Ricci scalar
curvature $R$, $g$ is the determinant of 4D metric $g_{\mu\nu}$
and $L_m$ is the matter Lagrangian. The line
element for the isotropic, homogeneous, and curvature free
Friedmann-Robertson-Walker (FRW) model of the universe is as
follows
\begin{equation}
    ds^2=dt^2-a^2(t)(dx^2+dy^2+dz^2),
\end{equation}
where $a(t)$ is the scale factor of the universe.
So for flat FRW model, $R=6(2H^2+\dot{H})$, where
$H=\frac{\dot{a}}{a}$ is the Hubble parameter. The field equation
corresponding to the above action is given by
\begin{equation}\label{2}
f_R(R)R_{\nu\mu}-\frac{1}{2}f(R)g_{\nu\mu}+(g_{\nu\mu}\Box-\triangledown_{\mu}
\triangledown_{\nu})f_R(R)=8\pi T_{\nu\mu},
\end{equation}
where $T_{\mu
\nu}=(\rho+p)u_{\mu}u_{\nu}+pg_{\mu\nu}$ is the energy-momentum
tensor. Here $\rho$ and $p$ are respectively the energy density
and pressure of the fluid and $u^{\mu}$ is the fluid four vector.

%From equation (\ref{2}), we can obtain the following field
%equations:
% \cite{56}
%\begin{equation}
%    \rho+\frac{1}{2}f(R)+3(\dot{H}+H^2)f_R(R,T)-3H\dot{R}f_{RR}(R)=0.
%\end{equation}
In the context of our work we are using the
isotropic and homogeneous FRW model and its evolutionary equation
to reconstruct the DBI-essence dark energy model.

\section{Overview of DBI-essence model}

The DBI-essence model starts from the following Lagrangian.
\cite{16}
\begin{equation}
    L=T(\phi)\sqrt{1-\frac{\dot\phi^2}{T(\phi)}}+V(\phi)-T(\phi).
\end{equation}
Here $T(\phi)$ is a Bulk Brane tension and $V(\phi)$ is the potential.

From the scalar field Lagrangian, the pressure and
energy density for DBI-essence model can be given as follows
\cite{16}
\begin{equation}\label{6}
    \rho_{\phi}=(\gamma-1)T(\phi)+V(\phi)
\end{equation}
and
\begin{equation}\label{7}
    p_{\phi}=(1-\frac{1}{\gamma})T(\phi)-V(\phi).
\end{equation}
The conservation equation for any scalar field always produce the
Klein-Gordon equation which gives us the dynamical nature of
scalar fields. In the case of the DBI-essence model, the modified
scalar field Klein-Gordan equation can be written as follows
\cite{16}
\begin{equation}
    \ddot{\phi}-\frac{3T'(\phi)}{2T(\phi)}\dot{\phi}^2+T'(\phi)
    +\frac{3}{\gamma^2}\frac{\dot{a}}{a}\dot{\phi}+\frac{1}{\gamma^3}(V'(\phi)-T'(\phi))=0.
\end{equation}
We assume $V(\phi)=T(\phi)=n\dot{\phi}^2$ as used in
\cite{33}, where  $\gamma$ is given by
$\gamma=\sqrt{1-\frac{\dot{\phi}^2}{T(\phi)}}$.\par Using this
assumption in equations (\ref{6}) and (\ref{7}) we get
\begin{equation}
    \frac{1}{2}\dot{\phi}^2=\frac{1}{2}\sqrt{\frac{n-1}{n}}[\rho_{\phi}+p_{\phi}].
\end{equation}
Rewriting the scalar field potential and tension in terms of
scalar field density and pressure, we obtain
\begin{equation}
     V(\phi)=T(\phi)=\sqrt{n(n-1)}[\rho_{\phi}+p_{\phi}].
\end{equation}

\section{Coupling between $f(R)$ gravity and DBI-essence DE and its reconstruction}

We proceed with the reconstruction of the DBI-essence model by
introducing coupling between the DBI-essence model and $f(R)$
model. In our study, modified gravity is the reconstructing
system, and the DBI-essence scalar field is reconstructed. The
coupled action can be written as follows \cite{40}
\begin{equation}\label{12}
S=\int{d^4
x\sqrt{-g}[f(R)+T(\phi)\sqrt{1-\frac{\dot{\phi}^2}{T(\phi)}}+V(\phi)-T(\phi)+L_m]}.
\end{equation}
The Field equations corresponding to the above action are as
follows
\begin{equation}
    3H^2=\frac{1}{f'(R)}[\rho_m+\rho_R+\rho_{\phi}]
\end{equation}
and
\begin{equation}
    3H^2+2\dot{H}=\frac{1}{f'(R)}[-p_m-p_R-p_{\phi}],
\end{equation}
where $\rho_R$ and $p_R$ represent the energy density and pressure
in context of $f(R)$ gravity. Their expressions are as follows
\begin{equation}\label{14}
    p_R=\frac{1}{2}[f(R)-Rf'(R)]+[2H\dot{R}+\ddot{R}]f''(R)+\dot{R}^2f'''(R)
\end{equation}
and
\begin{equation}\label{15}
    \rho_R=\frac{1}{2}[-f(R)+Rf'(R)]-3H\dot{R}f''(R).
\end{equation}
Here, prime denotes derivative w.r.t. $R$. The
energy density due to dark matter is given by
 \begin{equation}\label{16}
    \rho_m=\rho_{m0} a^{-3},
\end{equation}
where $\rho_{m0}$ is present value of the energy density.
Using equations (\ref{12})- (\ref{16}), the
expression of the reconstructed scalar field density and pressure
for the coupled system are as follows:
\begin{equation}
    \rho_{\phi}=3H^2f'(R)-\rho_m-\frac{1}{2}[-f(R)+Rf'(R)]+3H\dot{R}f''(R)
\end{equation}
and
\begin{equation}
    p_{\phi}=-(3H^2+2\dot{H})f'(R)-p_{m}-\frac{1}{2}[f(R)-Rf'(R)]-[2H\dot{R}+\ddot{R}]f''(R)-\dot{R}^2f'''(R).
\end{equation}

\section{Overview of different types of singularities and thermodynamic energy conditions}
The types of finite time singularities are given
as follows \cite{59,60,44}:
\begin{itemize}
    \item For any finite time, $t = t_s$ if we have $a\rightarrow\infty$; $\rho\rightarrow\infty$;
    $p\rightarrow\infty$,it is known as Type-I or Big Rip singularity.
    \item  For any finite time, $t = t_s$, if we have $a\rightarrow a_s$; $\rho\rightarrow \rho_s$;
    $p\rightarrow\infty$, it is Type-II or Sudden singularity.
    \item For any finite time, $t = t_s$ if we have $a\rightarrow a_s$; $\rho\rightarrow\infty$; $p\rightarrow\infty$, it is called Type III singularity. It is found in EOS of type$p =
    -\rho-A\rho^{\alpha}$.
    \item  At any finite time, $t = t_s$, if we have $a\rightarrow a_s$; $\rho\rightarrow \rho _s$; $p\rightarrow\infty$ and also the Hubble parameter and its first derivative remain finite, but its higher derivatives diverge, it is Type IV singularity. This kind of singularity is found in $p = - \rho -
    f(\rho)$.
\end{itemize}
The thermodynamic energy conditions can be derived from the
Raychaudhuri equation. For a congruence of time-like and null-like
geodesics, the Raychaudhuri equations are given in the following
forms that can be also found in the following
expression in \cite{59,60,22}\par
\begin{center}
$\frac{\mathrm{d} \theta}{\mathrm{d} \tau}=-\frac{1}{3}\theta^2-\sigma_{\mu\nu}\sigma^{\mu\nu}+w_{\mu\nu}w^{\mu\nu}-R_{\mu\nu}u^{\mu}u^{\nu}$
\end{center}
and
\begin{center}
$\frac{\mathrm{d} \theta}{\mathrm{d} \lambda}=-\frac{1}{3}\theta^2
-\sigma_{\mu\nu}\sigma^{\mu\nu}+w_{\mu\nu}w^{\mu\nu}-R_{\mu\nu}n^{\mu}n^{\nu}$,
\end{center}\par
where $\theta$ is the expansion factor, $n^{\mu}n^{\nu}$ is the
null vector, and $\sigma_{\mu\nu}\sigma^{\mu\nu}$ and
$w_{\mu\nu}w^{\mu\nu}$ are the shear and rotation associated with
the vector field $u^{\mu}u^{\nu}$. For attractive gravity we may
write followings \par
\begin{center}
$R_{\mu\nu}u^{\mu}u^{\nu}\geq0$ and
$R_{\mu\nu}n^{\mu}n^{\nu}\geq0$.
\end{center}\par
For our matter-fluid distribution, the energy conditions are as
follows:

\begin{itemize}
    \item [i.] Null energy condition (NEC)=  $\rho+p\geq0$.
    \item [ii.]  Weak energy condition (WEC)=$\rho\geq0$ and
    $\rho+p\geq0$.
    \item [iii.] Strong energy condition (SEC)= $\rho+3p\geq0$ and
    $\rho+p\geq0$.
    \item [iv.] Dominant energy condition (DEC)=$\rho\geq0$ and $-\rho\leq
    p\leq\rho$.
\end{itemize}

\section{Overview of black Hole and Wormhole Mass accretion}

We start with the study of mass accretion of black hole which is
derived in \cite{33}. The mass accretion of black holes can be
written as
\begin{equation}
    \dot{M}=-4\pi r^2 T_{0}^1.
\end{equation}
Here $T_{\mu \nu}$ is the energy momentum  tensor for the fluid
given by
 $T_{\mu \nu}=(\rho_{\phi}+\rho_{m}+p_{\phi}+p_m)u_{\mu}u_{\nu}-g_{\mu\nu}(p_{\phi}+p_m)$.
Therefore, the above equation becomes as follows
\begin{equation}
    \dot{M}=-4\pi A M^2 (\rho_{\phi}+\rho_{m}+p_{\phi}+p_m),
\end{equation}
where $A$ is a positive constant (details can be found in
\cite{38}). Similarly, the mass accretion of wormholes can be
written as follows \cite{33}
\begin{equation}
    \dot{M}=4\pi r^2 T_{0}^1.
\end{equation}
We can again re-write this equation using the functional form of
energy momentum tensor $T^{1}_0$ for wormhole, as follows
\begin{equation}
    \dot{M}=4\pi B M^2 (\rho_{\phi}+\rho_{m}+p_{\phi}+p_m),
\end{equation}
where $B$ is a positive constant which can be
found in \cite{44}. We will discuss the mass accretion with the
reconstructed scalar field energy density and pressure for
different scale factors($A$ and $B$ are the constants to
differentiate the black holes and Wormholes mass accretion). To
discuss the accretion phenomena, we consider pressureless dark
matter or cold dark matter i.e., $p_m=0$ in the subsequent
sections.

\section{Different scale factors and mass accretion analysis}

We will proceed with the discussion of the application of our
reconstructed system by considering four different types of scale
factor. We study the simplest form of $f(R)$ modified gravity
\cite{6} as
\begin{equation}
    f(R)=\lambda R^2,
\end{equation}
where $\lambda$ is constant and taken as $\lambda\in (I^-)$ for first two scale factor and $\lambda\in (I^+)$ for last two scale factors.

\subsection{Scale factor $a(t)=a_0(a_1+nt)^m$}
According to the results of section 4 with power-law form of scale
factor $a(t)=a_0(a_1+nt)^m$ \cite{31}, where $a_0,a_1,n,m$ are
positive constants, we can represent $\rho(\phi)$ and $p(\phi)$.\\
Here $a_1$ removes the initial singularity and $m>1$ for
accelerated universe. We plot $\rho(\phi)$ and $p(\phi)$ in Figs.
1 and 2. This $\rho(\phi)$ and $p(\phi)$ can be used in equations
(\ref{15}) and (\ref{16}) and the variations of kinetic energy and
potential energy can be observed in figures 3 and 4.

\begin{figure}[H]
\centering
\begin{minipage}[b]{0.4\textwidth}
    \includegraphics[width=\textwidth]{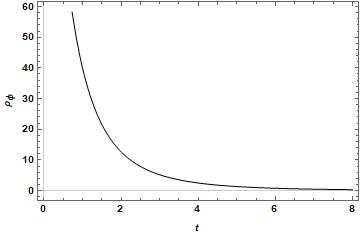}
    \caption{Plot of $\rho_{\phi}$ vs $t$ with $m = 2$, $n = 1/2$, $a_1 = a_0 = 1$ and $\lambda = -5$}
\end{minipage}
\hfill
\begin{minipage}[b]{0.4\textwidth}
    \includegraphics[width=\textwidth]{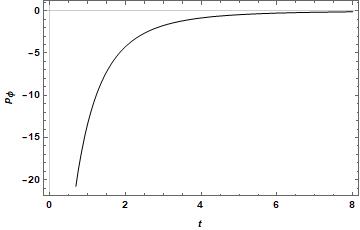}
    \caption{Plot of $p_{\phi}$ vs $t$ with $m = 2$, $n = 1/2$, $a_1 = a_0 = 1$ and $\lambda = -5$}
\end{minipage}
\end{figure}
\begin{figure}[H]
\centering
\begin{minipage}[b]{0.4\textwidth}
    \includegraphics[width=\textwidth]{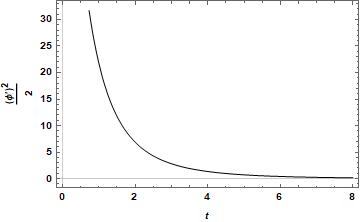}
    \caption{Plot of $\frac{1}{2}(\dot{\phi})^2$ vs $t$ with $m = 2$, $n = 1/2$, $a_1 = a_0 = 1$ and $\lambda = -5$}
\end{minipage}
\hfill
\begin{minipage}[b]{0.4\textwidth}
    \includegraphics[width=\textwidth]{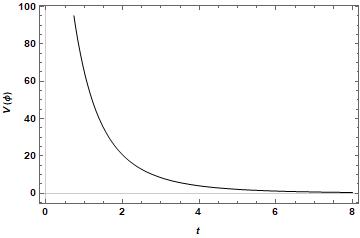}
    \caption{Plot of $V(\phi)$ vs $t$ with $m = 2$, $n = 1/2$, $a_1 = a_0 = 1$ and $\lambda = -5$}
\end{minipage}
\end{figure}

\subsubsection{Energy conditions with reconstructed DBI-essence model and effective system}

We will discuss the thermodynamic energy conditions w.r.t. the
scalar field energy density and pressure that are obtained for the
first assumed scale factor. The graphical representations for
energy conditions are shown in Figs. 5 to 7.
\begin{figure}[H]
\centering
\begin{minipage}[b]{0.25\textwidth}
    \includegraphics[width=\textwidth]{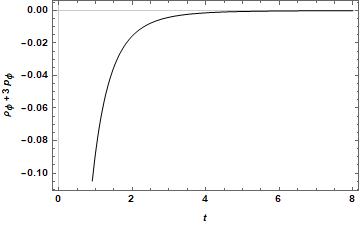}
    \caption{Plot of $\rho_{\phi}+3p_{\phi}$ with $t$ for $m = 2$, $n = 1/2$, $a_1 = a_0 = 1$ and $\lambda = -5$}
\end{minipage}
\hfill
\begin{minipage}[b]{0.25\textwidth}
    \includegraphics[width=\textwidth]{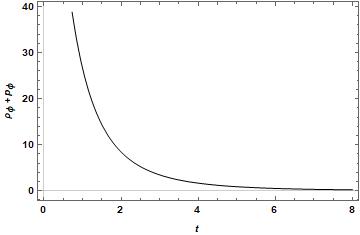}
    \caption{Plot of $\rho_{\phi}+p_{\phi}$ with $t$ for $m = 2$, $n = 1/2$, $a_1 = a_0 = 1$ and $\lambda = -5$}
\end{minipage}
\hfill
\begin{minipage}[b]{0.25\textwidth}
    \includegraphics[width=\textwidth]{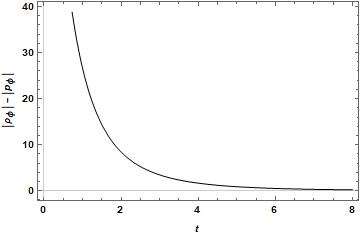}
    \caption{Plot of $\left |\rho_{\phi}\right |-\left |p_{\phi} \right |$ with $t$ for $m = 2$, $n = 1/2$, $a_1 = a_0 = 1$ and $\lambda = -5$}
\end{minipage}
\end{figure}

\subsubsection{Mass accretion formalism}

The basics of mass accretion has been discussed in section 6. The
mass accretion of black hole and wormhole is shown graphically in
in figs. 8 and 9.
\begin{figure}[H]
\centering
\begin{minipage}[b]{0.4\textwidth}
    \includegraphics[width=\textwidth]{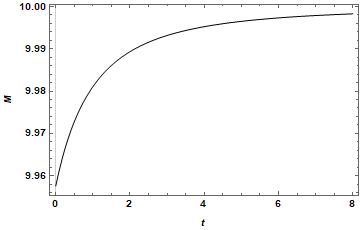}
    \caption{Plot of black hole mass $M(t)$ vs $t$ for $m = 2$, $n = 1/2$, $a_1 = a_0 = 1$ and $\lambda = -5$}
\end{minipage}
\hfill
\begin{minipage}[b]{0.4\textwidth}
    \includegraphics[width=\textwidth]{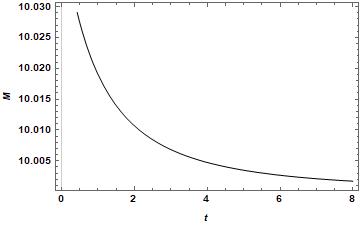}
    \caption{Plot of wormhole mass $M(t)$ vs $t$ for $m = 2$, $n = 1/2$, $a_1 = a_0 = 1$ and $\lambda = -5$}
\end{minipage}
\end{figure}

\subsection{Scale factor $a(t)=g+a_0(a_1+nt)^m$, $g$ = constant $> 0$ }

The second scale factor is of the form  $a(t)=g+a_0(a_1+nt)^m$,
where $m$ is even power. The importance of this constant $g$ is
that it resolves any past-time singularity. We can obtain the
expression $\rho(\phi)$ and $p(\phi)$ by the method discussed in
section 4. The variation of $\rho(\phi)$ and $p(\phi)$ can be
observed graphically in figures 10 and 11.
\begin{figure}[H]
\centering
\begin{minipage}[b]{0.4\textwidth}
    \includegraphics[width=\textwidth]{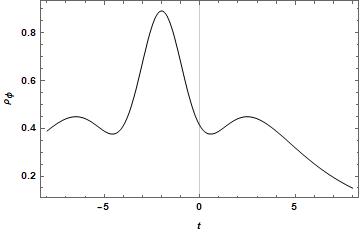}
    \caption{Plot of $\rho_{\phi}$ with $t$ for $m = 2$, $n = 1/2$, $a_1 = a_0 = 1$, $f = 5$ and $\lambda = -5$}
\end{minipage}
\hfill
\begin{minipage}[b]{0.4\textwidth}
    \includegraphics[width=\textwidth]{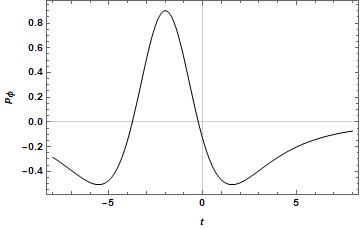}
    \caption{Plot of $p_{\phi}$ with $t$ for $m = 2$, $n = 1/2$, $a_1 = a_0 = 1$, $f = 5$ and $\lambda = -5$}
\end{minipage}
\end{figure}
The time variation of scalar field kinetic energy and potential
energy can be observed in Figs. 12 and 13.
\begin{figure}[H]
\centering
\begin{minipage}[b]{0.4\textwidth}
    \includegraphics[width=\textwidth]{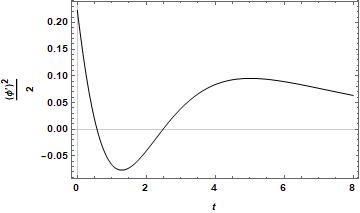}
    \caption{Plot of $\frac{1}{2}(\dot{\phi})^2$ with $t$ for $m = 2$, $n = 1/2$, $a_1 = a_0 = 1$, $f = 5$ and $\lambda = -5$}
\end{minipage}
\hfill
\begin{minipage}[b]{0.4\textwidth}
    \includegraphics[width=\textwidth]{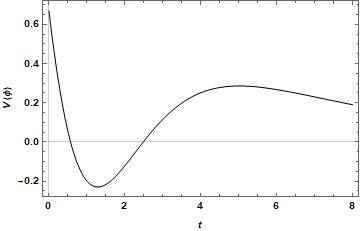}
    \caption{Plot of $V(\phi)$ with $t$ for $m = 2$, $n = 1/2$, $a_1 = a_0 = 1$, $f = 5$ and $\lambda = -5$}
\end{minipage}
\end{figure}

\subsubsection{Energy conditions with reconstructed DBI-essence model and effective system}

We study the thermodynamics energy conditions w.r.t. the scalar
field energy density and pressure shown graphically in Fig 10 and 11 in the previous
subsection.The graphical representations for energy conditions have
been shown below in Figs. 14 to 16.
\begin{figure}[H]
\centering
\begin{minipage}[b]{0.25\textwidth}
    \includegraphics[width=\textwidth]{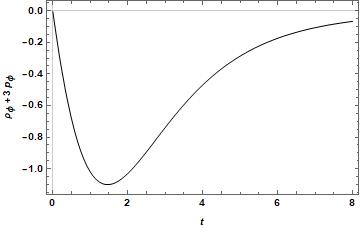}
    \caption{Plot of $\rho_{\phi}+3p_{\phi}$ with $t$ for $m = 2$, $n = 1/2$, $a_1 = a_0 = 1$, $f = 5$ and $\lambda = -5$}
\end{minipage}
\hfill
\begin{minipage}[b]{0.25\textwidth}
    \includegraphics[width=\textwidth]{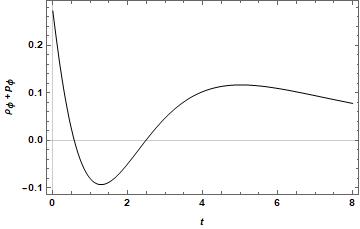}
    \caption{Plot of $\rho_{\phi}+p_{\phi}$ with $t$ for $m = 2$, $n = 1/2$, $a_1 = a_0 = 1$, $f = 5$ and $\lambda = -5$}
\end{minipage}
\hfill
\begin{minipage}[b]{0.25\textwidth}
    \includegraphics[width=\textwidth]{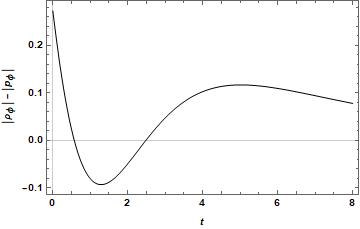}
    \caption{Plot of $\left |\rho_{\phi}\right |-\left |p_{\phi} \right |$ with $t$ for $m = 2$, $n = 1/2$, $a_1 = a_0 = 1$, $f = 5$ and $\lambda = -5$}
\end{minipage}
\end{figure}

\subsubsection{Mass accretion formalism}

The graphical representation of mass accretion of black hole and
wormhole is given in  Figs. 17 and 18.
\begin{figure}[H]
\centering
\begin{minipage}[b]{0.4\textwidth}
    \includegraphics[width=\textwidth]{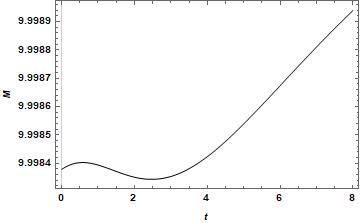}
    \caption{Plot of black hole mass accretion $M(t)$ vs $t$ with $t$ for $m = 2$, $n = 1/2$, $a_1 = a_0 = 1$, $f = 5$ and $\lambda = -5$}
\end{minipage}
\hfill
\begin{minipage}[b]{0.4\textwidth}
    \includegraphics[width=\textwidth]{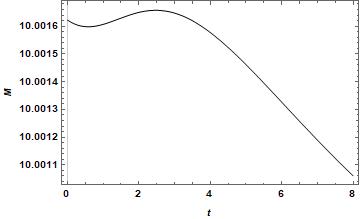}
    \caption{Plot of wormhole mass accretion $M(t)$ vs $t$ with $t$ for $m = 2$, $n = 1/2$, $a_1 = a_0 = 1$, $f = 5$ and $\lambda = -5$}
\end{minipage}
\end{figure}

\subsection{Scale factor $a(t)=a_0\exp{\alpha t^2}$}

The third assumed scale factor is of the form $a(t)=\exp{\alpha
t^2}$ as discussed in \cite{31}, we can derive $\rho(\phi)$ and
$p(\phi)$. The time variation of $\rho(\phi)$ and $p(\phi)$ can be
observed in Figs. 19 and 20.

The time variation of kinetic energy and potential energy for this
scale factor is shown in Figs. 21 and 22.
\begin{figure}[H]
\centering
\begin{minipage}[b]{0.4\textwidth}
    \includegraphics[width=\textwidth]{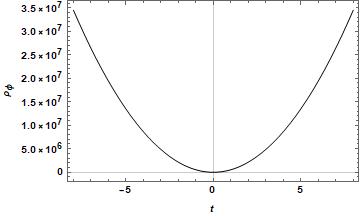}
    \caption{Plot of $\rho_{\phi}$ with $t$ for $\alpha = 5$, $m = 2$, $n = 1/2$, $a_1 = a_0 = 1$ and $\lambda = 5$}
\end{minipage}
\hfill
\begin{minipage}[b]{0.4\textwidth}
    \includegraphics[width=\textwidth]{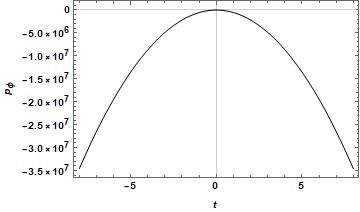}
    \caption{Plot of $p_{\phi}$ with $t$ for $\alpha = 5$, $m = 2$, $n = 1/2$, $a_1 = a_0 = 1$ and $\lambda = 5$}
\end{minipage}
\end{figure}
\begin{figure}[H]
\centering
\begin{minipage}[b]{0.4\textwidth}
    \includegraphics[width=\textwidth]{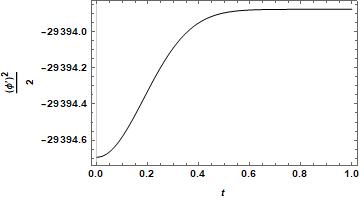}
    \caption{Plot of $\frac{1}{2}(\dot{\phi})^2$ with $t$ for $\alpha = 5$, $m = 2$, $n = 1/2$, $a_1 = a_0 = 1$ and $\lambda = 5$}
\end{minipage}
\hfill
\begin{minipage}[b]{0.4\textwidth}
    \includegraphics[width=\textwidth]{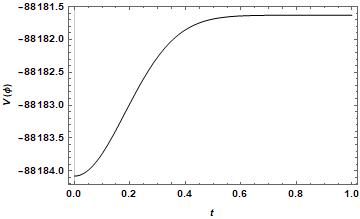}
    \caption{Plot of $V(\phi)$ with $t$ for $\alpha = 5$, $m = 2$, $n = 1/2$, $a_1 = a_0 = 1$ and $\lambda = 5$}
\end{minipage}
\end{figure}

\subsubsection{Energy conditions with reconstructed DBI-essence model and effective system}

The energy conditions corresponding the the above discussed
density and pressure are  graphically shown in Figs. 23 to 25.
\begin{figure}[H]
\centering
\begin{minipage}[b]{0.25\textwidth}
    \includegraphics[width=\textwidth]{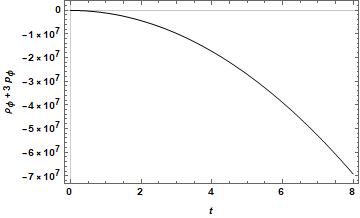}
    \caption{Plot of $\rho_{\phi}+3p_{\phi}$ with $t$ for $\alpha = 5$, $m = 2$, $n = 1/2$, $a_1 = a_0 = 1$ and $\lambda = 5$}
\end{minipage}
\hfill
\begin{minipage}[b]{0.25\textwidth}
    \includegraphics[width=\textwidth]{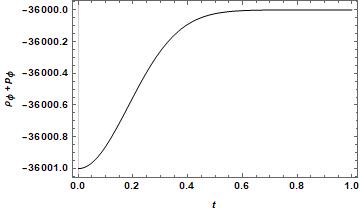}
    \caption{Plot of $\rho_{\phi}+p_{\phi}$ with $t$ for $\alpha = 5$, $m = 2$, $n = 1/2$, $a_1 = a_0 = 1$ and $\lambda = 5$}
\end{minipage}
\hfill
\begin{minipage}[b]{0.25\textwidth}
    \includegraphics[width=\textwidth]{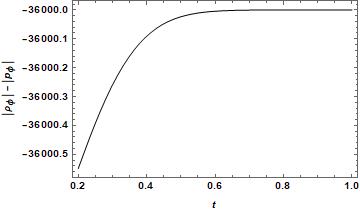}
    \caption{Plot of $\left |\rho_{\phi}\right |-\left |p_{\phi} \right |$ with $t$ for $\alpha = 5$, $m = 2$, $n = 1/2$, $a_1 = a_0 = 1$ and $\lambda = 5$}
\end{minipage}
\end{figure}

\subsubsection{Mass accretion formalism}

The mass accretion for the third type of scale factor is given  in
Figs. 26 and 27.
\begin{figure}[H]
\centering
\begin{minipage}[b]{0.4\textwidth}
    \includegraphics[width=\textwidth]{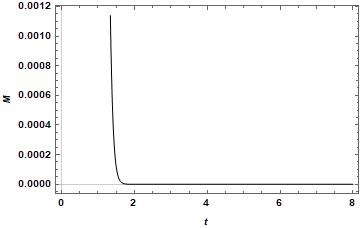}
    \caption{Plot of black hole mass $M(t)$ vs $t$ with $t$ for $\alpha = 5$, $m = 2$, $n = 1/2$, $a_1 = a_0 = 1$ and $\lambda = 5$}
\end{minipage}
\hfill
\begin{minipage}[b]{0.4\textwidth}
    \includegraphics[width=\textwidth]{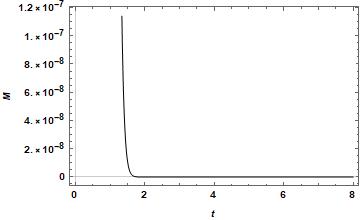}
    \caption{Plot of wormhole mass $M(t)$ vs $t$ with $t$ for $\alpha = 5$, $m = 2$, $n = 1/2$, $a_1 = a_0 = 1$ and $\lambda = 5$}
\end{minipage}
\end{figure}

\subsection{Scale factor $a(t)=a_0(\exp{\alpha t^2}+\exp{\alpha^2 t^4})$}

The fourth assumed scale factor is $a(t)=a_0(\exp{\alpha
t^2}+\exp{\alpha^2 t^4})$ as discussed in \cite{31}, we can derive
$\rho(\phi)$ and $p(\phi)$. The time variation of $\rho(\phi)$ and
$p(\phi)$ is shown in Figs. 28 and 29.

\begin{figure}[H]
\centering
\begin{minipage}[b]{0.4\textwidth}
    \includegraphics[width=\textwidth]{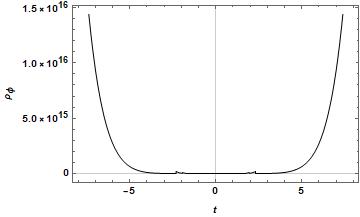}
    \caption{Plot of $\rho_{\phi}$ with $t$ for $\alpha = 5$, $m = 2$, $n = 1/2$, $a_1 = a_0 = 1$ and $\lambda = 5$}
\end{minipage}
\hfill
\begin{minipage}[b]{0.4\textwidth}
    \includegraphics[width=\textwidth]{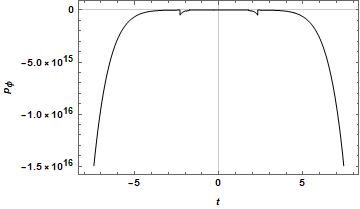}
    \caption{Plot of $p_{\phi}$ with $t$ for $\alpha = 5$, $m = 2$, $n = 1/2$, $a_1 = a_0 = 1$ and $\lambda = 5$}
\end{minipage}
\end{figure}

The time variation of kinetic energy and potential energy is shown
in Figs. 30 and 31.
\begin{figure}[H]
\centering
\begin{minipage}[b]{0.4\textwidth}
    \includegraphics[width=\textwidth]{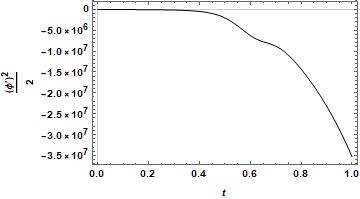}
    \caption{Plot of $\frac{1}{2}(\dot{\phi})^2$ with $t$ for $\alpha = 5$, $m = 2$, $n = 1/2$, $a_1 = a_0 = 1$ and $\lambda = 5$}
\end{minipage}
\hfill
\begin{minipage}[b]{0.4\textwidth}
    \includegraphics[width=\textwidth]{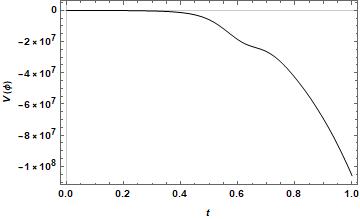}
    \caption{Plot of $V(\phi)$ with $t$ for $\alpha = 5$, $m = 2$, $n = 1/2$, $a_1 = a_0 = 1$ and $\lambda = 5$}
\end{minipage}
\end{figure}

\subsubsection{Energy conditions with reconstructed DBI-essence model and effective system}

The graphical representations of the energy conditions
corresponding to the above assumed scale factor are Figs. 32 to
34.
\begin{figure}[H]
\centering
\begin{minipage}[b]{0.25\textwidth}
    \includegraphics[width=\textwidth]{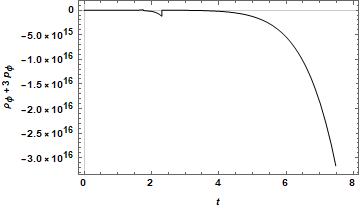}
    \caption{Plot of $\rho_{\phi}+3p_{\phi}$ with $t$ for $\alpha = 5$, $m = 2$, $n = 1/2$, $a_1 = a_0 = 1$ and $\lambda = 5$}
\end{minipage}
\hfill
\begin{minipage}[b]{0.25\textwidth}
    \includegraphics[width=\textwidth]{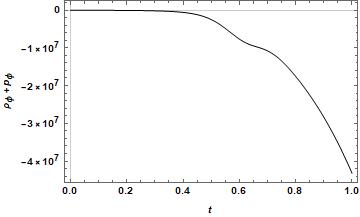}
    \caption{Plot of $\rho_{\phi}+p_{\phi}$ with $t$ for $\alpha = 5$, $m = 2$, $n = 1/2$, $a_1 = a_0 = 1$ and $\lambda = 5$}
\end{minipage}
\hfill
\begin{minipage}[b]{0.25\textwidth}
    \includegraphics[width=\textwidth]{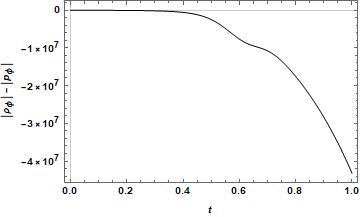}
    \caption{Plot of $\left |\rho_{\phi}\right |-\left |p_{\phi} \right |$ with $t$ for $\alpha = 5$, $m = 2$, $n = 1/2$, $a_1 = a_0 = 1$ and $\lambda = 5$}
\end{minipage}
\end{figure}

\subsubsection{Mass accretion formalism}

The variation of mass accretion of black hole and wormhole are
shown in  Figs. 35 and 36.
\begin{figure}[H]
\centering
\begin{minipage}[b]{0.4\textwidth}
    \includegraphics[width=\textwidth]{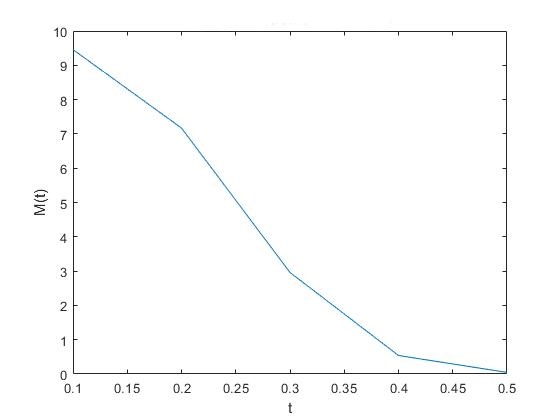}
    \caption{Plot of black hole mass $M(t)$ vs $t$ with $t$ for $\alpha = 5$, $m = 2$, $n = 1/2$, $a_1 = a_0 = 1$ and $\lambda = 5$}
\end{minipage}
\hfill
\begin{minipage}[b]{0.4\textwidth}
    \includegraphics[width=\textwidth]{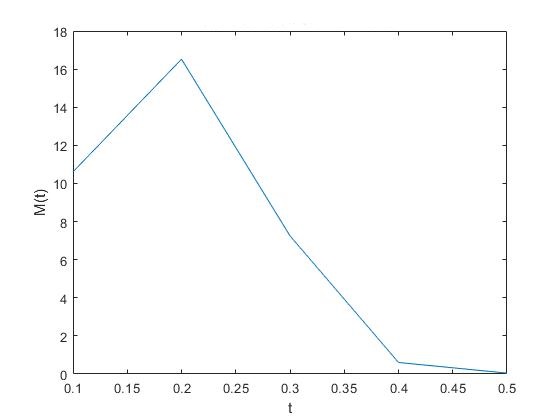}
    \caption{Plot of wormhole mass $M(t)$ vs $t$ with $t$ for $\alpha = 5$, $m = 2$, $n = 1/2$, $a_1 = a_0 = 1$ and $\lambda = 5$}
\end{minipage}
\end{figure}

\section{Removal of singularities}

In section 5, we have discussed the various types of singularities
that can be present in any model. In this paper, we have studied
four types of scale factors. It is observed that the scalar field
density and pressures of all the four types of scale factors are
continuous and finite and hence there is no singularity at an
initial time as well as at a future time. It is also observed that
for the first scale factor in the time range $t>0$ and for all
other scale factors in the time range $t\in(-\infty,\infty)$ we
did not find any kind of singularities. Thus from fluid and
geometry parameters, we can conclude that our model is free of any
singularity.

\section{Physical Analysis and Discussions}

In this work, we have discussed four types of
scale factors, along with their pictorial representation of
reconstructed scalar field energy densities, pressures, kinetic
energies, and potential energies. Later, we discussed energy
conditions and mass accretion with respect to the reconstructed
scalar field of black holes and wormholes, respectively. The
detailed discussions of those results have been given in the
proceeding paragraphs.\par The energy densities
have similar nature for the third and fourth type scale factors.
Both of them have increasing nature after the point of cosmic
bounce. The first type scale factor provides decreasing type
energy density, whereas the second type bouncing scale factor
gives some variation in nature with time. For the second type
scale factor, the energy density firstly decreases with time, then
after cosmic bounce, it increases with time and then again
decreases. The turning points provide some phase transition whose
details discussion is beyond the scope of that work.\par
The pictorial representations of the pressures
provide exactly opposite nature of the energy densities of them.
This does not happen for the second type scale factor. For the
second type, the energy density and pressure have similar nature.
Except for the second kind scale factor, the pressures of all
other scale factors have negative values, and hence we may
conclude except for the second type scale factor, all others can
represent just the accelerated expansion of the universe. The
pressure representation of the second scale factor can conclude
that we can have a deceleration phase also somewhere after the
bounce and late-time acceleration.\par The nature
of scalar field kinetic energy and potential energy changes with
changing the scale factors, but they are surprisingly evolving in
similar ways when we analyze them with fixed type scale factor.
The potential of the first type scale factor provides the idea of
slow roll variation. The potential for second type scale factor
can provide negative results, which also proves the existence of
an attractive gravity-dominated universe for some time after the
cosmic bounce.\par The mass accretion rate
decreases for both the inflationary and bouncing type scale
factors, i.e., for third and fourth type scale factors. But for
first and second type scale factors, the mass accretion rate of
the black hole is increasing, and for the wormhole, it is
decreasing in nature.\par In the analysis of
energy conditions, we can find that the strong energy conditions
are mostly violated. The other energy conditions are also violated
except for the first and second type scale factors.\par
Thus, we have found a comparison between the
results we found for four types of scale factors with
reconstructed Dirac-Born-Infeld scalar field model with modified
$f(R)=\lambda R^2$ gravity.

\section{Concluding remarks}

In this paper, we have discussed the reconstructed scalar field
theory in terms of the DBI-essence model. The reconstruction has
been done by introducing coupling between the DBI-essence scalar
field dynamic dark energy model and modified gravity, particularly
$f(R)$ gravity. We have observed late-time decreasing nature of
reconstructed energy density for the first two power-law scale
factors whereas an increasing nature for the last two exponential
scale factors. The pressures for the first two scale factors have
increasing nature and for the last two have to decrease
nature.\par In the analysis of mass accretion, we
have found different natures for Black holes and Wormholes mass
accretion for different scale factors. The variation of mass
accretion for the inflationary cosmology remains similar in
nature, irrespective of the nature of the scale factor.\par
The last two scale factors have been taken to
discuss the bouncing universe and inflationary phase at the same
time. The second scale factor is our assumption on power-law
cosmology to resolve the initial time singularity at $t\rightarrow
0$ as well as $t\rightarrow 0^{-}$ or $t < 0$. The first scale
factor also resolves the initial singularity but only at
$t\rightarrow 0$.\par Overall, we have discussed
the reconstruction mechanism and mass accretion with the
discussion of energy conditions in some cosmic phases at
primordial times of the universe.

\section*{Limitations of this work}

The first two scale factors have been assumed, and the last two
have been taken from works of Bamba et al. \cite{31}. We have
investigated the results using reconstructed scalar field theory
but didn't use any theoretical analysis on mass accretion.

\section*{Acknowledgement}
The authors are thankful to the editors and
reviewers for their insightful comments for valuable improvements
to our paper.

\end{document}